\documentclass[useAMS]{mn2e}
\usepackage{times}
\usepackage{amssymb}

\def\simlt{\lower.5ex\hbox{$\; \buildrel < \over \sim \;$}}
\def\simgt{\lower.5ex\hbox{$\; \buildrel > \over \sim \;$}}
\def\simpropto{\lower.2ex\hbox{$\; \buildrel \propto \over \sim \;$}}

\newcommand{\be}{\begin{equation}}
\newcommand{\ee}{\end{equation}}
\newcommand{\bea}{\begin{eqnarray}}
\newcommand{\eea}{\end{eqnarray}}
\newcommand{\nn}{\nonumber}
\newcommand{\beqs}{\begin{subeqnarray}}
\newcommand{\eeqs}{\end{subeqnarray}}
\newcommand{\pd}{\partial}

\def\grad{\vec{\nabla}}

\def\rot{\vec{\nabla}\times}

\title[First Stars] {\Large \bf  On the First Generation of Stars}

\author[Silk \& Langer]
{Joseph Silk and Mathieu Langer\\Astrophysics, University of Oxford\\Denys Wilkinson Building, 
Keble Road, Oxford OX1 3RH, United Kingdom}

\begin{document}

\date{Draft version \today}

\pagerange{\pageref{firstpage}--\pageref{lastpage}} \pubyear{2005}

\maketitle

\label{firstpage}

\begin{abstract}
We argue that the first
stars may have spanned the conventional mass range
 rather than be identified with the Very
Massive Objects ($\sim 100-10^3 \rm M_\odot$) favoured by numerical
simulations. Specifically, we find that magnetic field generation processes
acting in the first protostellar systems suffice to produce fields that  exceed 
the threshold for MRI instability to operate and thereby allow the MRI dynamo 
to generate equipartition-amplitude magnetic fields on protostellar mass
scales below $\sim 50 \rm M_\odot$. Such fields allow primordial star
formation to occur at essentially any metallicity by regulating angular
momentum transfer, fragmentation, accretion and feedback in much the same way
as occurs in conventional molecular clouds.
\end{abstract}


\section{Introduction}

The formation of the first stars should in principle be simpler to
understand than present-day star formation.  The usual prediction that
the mass of the first stars ${\mathcal{O}}(100-1000)\rm M_\odot$ is
attributable to dominance of H$_2$ cooling and consequent high
temperature and accretion rate (Abel, Bryan and Norman 2002; Bromm,
Coppi and Larson 2002).  It is appropriate to question this result,
that only very massive stars formed, for at least four
phenomenological reasons.

Firstly, one finds solar mass stars at $\rm [Fe] \simlt -4$.  These
may be contaminated by companion stars that exploded as hypernovae,
whose ejecta are depleted in Fe relative to N and C and this
interpretation is consistent with the measured abundance ratios for a
handful of extreme metal-poor stars. However not all metal poor halo
stars display such anomalies, and a substantial fraction at the
$10^{-4}$ solar level have abundance ratios that are consistent with
conventional SNII precursors. There are at least two halo stars known
at $\rm [Fe] \simlt -5$.  HE 0107-5240 is a giant with [Fe/H] =-5.3,
but enhanced nitrogen, carbon and oxygen: [N/H]= -3.0, [C/H]= -1.3 and
[O/H]=-2.9 (Christlieb et al. 2004).  HE 1327-2326 is a main sequence
star (or subgiant) with an iron abundance about a factor of 2 lower
than HE 0107-5240.  In this latter case, both nitrogen and carbon are
enhanced relative to iron by about 4 dex, while there is only a
comparable upper limit on oxygen (Frebel et al. 2005).

Appeal to preenrichment by a core collapse $\sim 25 \rm\,M_\odot$
supernova of Population III abundance with fallback fits the abundance
patterns well (Umeda and Nomoto 2003), although oxygen may possibly be
overproduced in the case of HE 0107-5240 (Bessell, Christlieb and
Gustafsson 2004).  The example of HE 1327-2326 eliminates any internal
enrichment source (e.g. convective dredge-up), nor is there evidence
for a binary companion in the case of HE 0107-524, thereby also
suggesting that mass transfer from an AGB star is an unlikely
explanation. The high CNO abundances argue for a common explanation
involving enrichment of the primordial cloud by Type II supernovae of
primordial abundance. Indeed, the abundance patterns in extremely
metal poor halo stars suggest that enrichment was produced by
Population III stars in the mass range 20-130 $M_\odot$ (Umeda and
Nomoto 2005).

Secondly, the broad emission line regions of very high redshift
quasars, reveal high elemental abundances that appear to have also
been generated by conventional SNII precursors.  In particular, the
nearly constant FeII/MgII emission line ratios over $0<z<5$ requires
intense SNII activity at a redshift as high as 9 (Dietrich, Hamann,
Appenzeller and Vestergaard 2003).

Thirdly, the chemical yield predictions from primordial very massive
stars, when normalised to the inferred ionising photon output required
to reionise the universe at high redshift, do not correspond to
observed abundances in any primitive environments.  In particular, the
pair-instability supernova nucleosynthetic signatures generated by
stars with initial masses in the range 130 to 260 M$_\odot$ are not
seen either in the intergalactic medium, including both Lyman alpha
forest and damped Lyman alpha absorption systems, or in extremely
metal-poor halo stars (Daigne et al. 2005).

Finally, from the theoretical perspective with regard to cooling, the
primordial metal abundance pattern has profound consequences for the
thermal balance and chemical composition of the gas, and hence for the
equation of state of the parental cloud.  Spaans and Silk (2005) find
that the polytropic index is soft for low oxygen abundance
enhancements, [O/H]$<-3$, as appropriate for Population III, but
stiffens to a polytropic index $\gamma$ larger than unity for
[O/H]$>10^{-2}$ due to the large opacity in the CO and H$_2$O cooling
lines.  Hence Pop III star formation is efficient, especially before
hypernova enrichment and associated oxygen enhancement has occurred.
There should be no obstacle to forming stars over a wide range of
masses even in the absence of significant fine-structure cooling. On
the contrary, once the polytropic index stiffens, at [O/H]$\simgt -2$,
the IMF should change and, specifically, flatten, as argued by Spaans
and Silk (2000).  Star formation is subsequently less efficient, in
part because of massive star feedback in addition to the stiffening of
the equation of state, as is required in most discussions of galaxy
formation in order to avoid premature consumption of the gas
reservoir.

Of course, to do justice to the very massive primordial star
hypothesis, the typical expectation of one very massive star per
primordial cloud is unlikely to produce enough metals to raise the
mean IGM metallicity to a value of $\sim 10^{-4}-10^{-2.5}$ solar, as
observed in the Lyman alpha forest (Schaye et al. 2003) and as also
required to significantly modify the cooling and thereby enhance
fragmentation.  This is especially plausible as the ejecta are mostly
not retained in the shallow potential wells of the pregalactic clouds
(Norman, O'Shea and Paschos 2004).  Rather, one has the impression
that the first stars that produced metals in significant amounts, in
particular as monitored in the abundance patterns of the Lyman alpha
forest at modest overdensity and in halo stars at $\rm [Fe] \simlt
-4,$ spanned the conventional stellar mass range.

A more fundamental issue may be that the elegant state of the art
simulations with cosmological initial conditions have nothing directly
to say about the primordial stellar initial mass function, nor for
that matter about the physics underpinning the inferred transfer of
angular momentum, but only about the mass function of gas clumps. The
ultimate fate of these gas clumps has not yet been convincingly
demonstrated.

Notice that recent simulations by Saigo, Matsumoto and Umemura (2004)
show that Population III stars are prone to form in binary systems
through the formation of a rotationally supported disk. For that to
happen, the initial angular momentum of the parent cloud needs to be
small, the centrifugal force being at most $10\% - 30 \%$ of the
pressure force. However, the accretion efficiency seems not affected
by the binary formation, and the binary Pop III stars are again
predicted to be very massive.

Of course, all of this begs the question as to whether a softened EOS
can lead to low or intermediate mass star formation in situations of
very low metallicity, or at any rate, reduce the characteristic
stellar mass to below the lower limit for pair instability SN.
Certainly, the clouds from which stars with [Fe/H] $\simlt -5$ formed
seem to have had a soft equation of state.  The minimum fragment mass
is lowered at [O/H]$\simgt -3$ due to enhanced fine-structure cooling
(Bromm and Loeb 2003), but it is likely that turbulent fragmentation
(Klessen et al. 2004) enhanced by the softening of the equation of
state is the key to low-mass fragmentation, for example via disk
formation. According to this viewpoint, it is the equation of state in
the initial collapsing cloud that controls the IMF.

An alternative view is that the final fragment mass remains high, as
suggested by the numerical simulations, but that a second mode of
fragmentation associated with the onset of dust cooling at very high
density may lead to low mass star formation (Omukai et al. 2005).  The
latter study finds that a metallicity of $Z\simgt 10^{-5}Z_\odot$ at a
density in excess of $10^{10}\rm cm^{-3}$ marks this transition.
However Spaans and Silk (2005) argue that the gas properties, and in
particular the equation of state, at much lower densities, typically
the range associated with saturation of H$_2$ cooling at $\sim 10^4\rm
cm^{-3}$, most likely determine the IMF.

The purpose of this paper is to demonstrate that a plausible
alternative to dust cooling in primordial clouds involves recourse to
a mechanism that is universally accepted as being an essential key to
controlling the IMF in conventional molecular clouds at the onset of
collapse, namely magnetically-regulated fragmentation and accretion.
The essence of our argument will be to derive the magnetic field
strength in primordial star formation.  An important conclusion is
that one should continue to find low mass, and especially abundance
signatures of intermediate mass, stars down to essentially zero
metallicity. Note that the possible effects of magnetic fields on
  primordial star formation have also been investigated in a pioneering paper
by Tan and
  Blackman (2004), and more recently by Machida et al. (2006).

\section{The transition from primordial star formation}

For the moment, pending higher resolution simulations with adequate
exploration of the EOS, it may be grasping at straws to argue that Pop
III could contain a substantial or even a dominant low mass component.
Nevertheless, an analytic exploration is warranted. In this note,
therefore, an alternative approach is addressed.  A reasonable
hypothesis, prevalent in the star formation community, is that
magnetic fields are essential to the generation of the initial stellar
mass function in conventional star-forming clouds  (e.g. Mestel,
  1965; Lizano and Shu, 1989; Shu et al., 2004), even if precise
  details of the actual regulation mechanism still remain to be worked
  out. If this is indeed the case, one may then ask what happened in
the first clouds, when presumably magnetic fields were absent at any
significant dynamical level? We refer to this as Primordial Star
Formation.  And was there a seamless transition to Current Epoch Star
Formation, involving no drastic change in the underlying physics?

One common, and generally accepted, viewpoint among cosmologists is
that metallicity was a primary driver. When the metallicity was less
than about 0.001 of the solar value, the dominant coolants were $H$
atoms and H$_2$ molecules. These sufficed only to cool to about 200K.
The inefficient cooling is the primary reason that both analytic and
numerical calculations of the minimum fragmentation mass in primordial
clouds yields high accretion rates onto protostellar cores
$(\dot{M}_\star\approx c_s^3/G)$ that result in characteristic
primordial stellar masses of $\sim 100- 1000 \rm M_\odot.$

However the protostellar core is only $0.01\rm M_\odot$ and grows by
accretion.  One may try to understand how low mass stars could form in
a primordial, or nearly primordial, environment if fragmentation,
itself inefficient, was aided by magnetic feedback.  Consider the
following hypothesis. Magnetic fields are generated by a dynamo
associated with magneto-rotational instabilities at work in
protostellar or circumstellar disks.  There are several necessary
conditions for such an MRI dynamo to operate.  Firstly, the proton
Larmor radius must be less than the disk scale-height.  Secondly,
there are important conditions imposed on the wavelength and on the
growth rate of the MRI instability.  Thirdly, there must be a seed
field that exceeds the minimal field under which the MRI dynamo
operates.  In the following, we will examine each one of these issues,
within the framework of a simple steady state circumstellar disk
model. This will allow us to obtain a critical mass of the central
(proto-)star above which all the conditions are fullfilled for the MRI
to work.

\subsection{General equations}
We start by considering a three fluid model, including free electrons,
ions (protons, for simplicity) and neutral gas, of
respective number densities $n_e, n_i$ and $n_n$. Setting $n \equiv
n_e = n_i$ global charge neutrality implies that the electric current
is $\vec{j} = - e n \vec{V}$, where $\vec{V} = \vec{v}_e - \vec{v}_i$
is the electron velocity relative to ions.  Similarly to the procedure
followed by Mestel (2003), we can combine the equation of motion of
electrons with that of ions to obtain the following generalized Ohm's
law equation,
\bea
\vec{E} &=& - \frac{\vec{v}\times\vec{B}}{c} - \frac{\grad p_e}{n_e} +
 \left(\kappa_e + \kappa - \frac{\kappa_e^2}{\kappa_e +
     \kappa_i}\right)\frac{B}{enc}\vec{j} \nonumber\\
 & &+ \left(1 - 2 F  \frac{\kappa_e}{\kappa_e + \kappa_i} \right)
 \frac{\vec{j}\times\vec{B}}{enc} + \frac{\kappa_e}{\kappa_e+ \kappa_i}\frac{F}{en}\left(1-f
   F^{-1}\right) \grad p \nonumber\\
& &+ \frac{ F^2}{enB\kappa_i}\left[\left(1-f F^{-1}\right)\grad p
   -\frac{\vec{j}\times\vec{B}}{c}\right]\times\vec{B} \label{ohms}
\eea
where $\kappa^{-1}= \omega \tau, \kappa^{-1}_e= \omega \tau_e$ and
$\kappa^{-1}_i= \omega \tau_i$ measure how strongly the charged
particles are coupled to the magnetic field with respect to their
collisional interactions ($\omega = eB/m_ec$ and $\omega_i = eB/m_ic$
are the electron and ion gyrofrequencies respectively, $\tau$ being
the mean electron-ion collision time, and $\tau_e$ and $\tau_i$ the
mean times for electron and ion collisions with neutrals). Note that
the products $\kappa_qB$ are independent of the magnetic field
amplitude $B$. The density fraction of neutral particles is given by
$F = \rho_n/(\rho_i + \rho_n)$, and their fractional contribution to
the total pressure $p$ is  $f = p_n/p$. The rest of the symbols
used above bear their usual meanings.

We get the magnetic field induction equation by taking the curl of
Eq. (\ref{ohms}). Ignoring the pressure terms, we obtain
\bea \pd_t \vec{B} 
&=& \rot \left[\left(\vec{v}+\vec{\Omega}\times\vec{r}\right)\times\vec{B}\right]
\nonumber \\
& &- \rot \left[\left(\kappa_e + \kappa - \frac{\kappa_e^2}{\kappa_e +
    \kappa_i}\right)\frac{B}{ne}\,\vec{j}\right.\nonumber\\
& &- \left(1 - 2 F \frac{\kappa_e}{\kappa_e + \kappa_i} \right)
\left(\frac{\vec{j}\times\vec{B}}{en}\right)\nonumber \\
& &+ \left. \frac{
  F^2}{eB\kappa_i}\left(\frac{\left(\vec{j}\times\vec{B}\right)\times\vec{B}}{n}\right)\right]
\eea 
in a local Keplerian frame corotating with the disk at angular
frequency $\Omega$. The second, third and fourth terms on the right
hand-side are the resistive, Hall and ambipolar diffusion terms
respectively. It is straightforward to check that we recover the usual
induction equation in the limit of vanishing density of neutrals ($F,f
\rightarrow 0$, and $\kappa_{e,i} \rightarrow 0$ since
$\tau_{e,i}\rightarrow\infty$).

Similarly, the equation of motion of the whole plasma is 
\be
\rho \frac{d\vec{v}}{dt} = \rho \grad \phi +
\frac{\vec{j}\times\vec{B}}{c} + \vec{F}_{\nu}  - \rho \left(2
  \vec{\Omega}\times \vec{v} + \vec{\Omega}\times\left(\vec{\Omega}\times\vec{r}\right)\right)
\ee
where $\vec{F}_{\nu}$ accounts for the viscous forces, and $d\cdot/dt
= \pd_t\cdot + \vec{\Omega}\times\cdot + \vec{v}\cdot\grad\cdot $ is
the full time derivative. Let us define the velocity departure from
Keplerian motion by $\delta \vec{v} = \vec{v} - \Omega
r\vec{e}_{\theta}$. The equation of motion then reads
\bea
\left[\pd_t + \Omega\pd_{\theta} + \delta\vec{v}\cdot\grad\right]\delta\vec{v} &=& - 2\Omega (\delta
v_r\vec{e}_\theta - \delta v_\theta \vec{e}_r) + 3 \Omega^2 r\vec{e}_r
\nonumber \\
& &+ \rho \grad \phi + \frac{\vec{j}\times\vec{B}}{c} + \vec{F}_\nu,\label{veloc}
\eea
and the induction equation simply becomes
\bea
\pd_t \vec{B} 
&=& \rot \left[\left(\delta\vec{v}+\Omega r\vec{e}_\theta\right)\times\vec{B}\right]\nonumber\\
& &- \rot \left[\left(\kappa_e + \kappa - \frac{\kappa_e^2}{\kappa_e +
    \kappa_i}\right)\frac{B}{ne}\,\vec{j}\right],\label{induct}
\eea
where we omitted the Hall and the ambipolar terms which are of second
order for initially weak magnetic fields.

Circumstellar disks are prone to gravitational instabilty. Especially
in the context of Pop III star formation, Tan and Blackman pointed out
that proto-stellar disks may develop gravitationally driven
turbulence, amenable to the $\alpha$ formalism (e.g. Gammie, 2001),
that provides angular momentum transport. As shown by Eqs.
(\ref{veloc}) and (\ref{induct}), gravitational instability may
amplify velocity fluctuations, which in turn can, in principle, drive
the growth of magnetic fields. 
The importance of gravito-turbulent effects 
 depends on the ratio of the first term and the last term of the right hand
 side of Eq. (\ref{induct}), which yields the actual magnetic Reynolds
 number. 
The factor multiplying the current on the
right hand side of Eq. (\ref{induct}) is
\be\label{factor}
\left[\kappa_e + \kappa - \frac{\kappa_e^2}{\kappa_e +
    \kappa_i}\right]\frac{B}{ne} = \frac{m_e c}{n\tau e^2}\left[1 +
  \frac{\tau}{\tau_e} \left(1+ \frac{m_e}{m_i} \frac{\tau_i}{\tau_e} \right)^{-1}\right]
\ee
Taking the mean collision time of charged particles with neutrals
from Draine et al. (1983), we obtain 
\be
\frac{\tau_i}{\tau_e}\sim 52\left(\frac{T}{10^4 \rm K}\right)^{1/2}
\ee
and 
\be
\frac{\tau}{\tau_e}\sim 1.66\times 10^{-3}\frac{n_n}{n_i}\left(\frac{T}{10^4 \rm K}\right)^{2},
\ee
and the magnetic Reynolds number is then
\begin{equation}
\mathcal{R}_M = \frac{\delta v r_{\rm grad}}{\eta [\,]}
\end{equation}
where the square brackets stand for the last term in equation (\ref{factor}),
accounting for the effects of neutrals. The presence of neutral particles in 
the disk would
enhance the dissipation effects. For initially weak magnetic fields,
this would  reduce the capability of gravitational instability
to amplify any weak magnetic seed fields, even if velocity fluctuations would be
amplified. 
Taking the velocity fluctuations to be of the
order of the sound speed in the disk, and the gradient scale being at most of
the order the disk radius, the magnetic Reynolds number reaches $\sim 
10^{14}$ at $T\sim 10^4$ K.

Based on the high Reynolds number, Tan and Blackman (2004) argue that local
gravitational instability develops, leading to an alpha-driven dynamo. This 
could provide a
mean of amplifying any initial seed fields to have a dynamical effect on
primordial star formation. However, recent simulations of massive 
self-gravitating disks find that
global gravitational instability modes dominate the energy transport (Lodato
and Rice 2005). If this is correct, as Balbus and Papaloizou (1999) initially
suggested, the alpha formalism would not be effective in field amplification.

Amplification of magnetic fields may however be driven by the action
of MRI turbulence. The role of neutrals, Hall effect and ambipolar
diffusion in weakly ionized disks has been investigated in several
papers (e.g. Wardle 1999, Salmeron and Wardle 2003, Kunz and Balbus
2004). However, as we discuss below, one does not expect proto-stellar
disks around forming Pop III stars to be cool enough so as to allow a
dominant amount of neutral particles. In the following, we therefore
consider a two fluid disk model only.

\subsection{Simple disk model}
To demonstrate the potential role of magnetic fields in Pop III
accretion, we use a very simple accretion disk model. We consider a
thin, nearly Keplerian disk in the steady state. In that case, the
accretion rate $\dot{M}$ is constant throughout the disk, and the
surface density
\be \Sigma(r) = \int_{-\infty}^{+\infty}\rho dz \simeq
2 H(r)\rho_0(r),\label{surf} \ee can be related to the accretion rate
via \be \nu \Sigma = \frac{1}{3 \pi} \dot{M}\left[1-\beta
  \left(\frac{r_i}{r}\right)^{1/2}\right]\label{accrate} 
\ee 
where $\rho (\rho_0)$ is the disk (midplane) gas density, $H$ is the
disk height scale, $\nu$ is the viscosity, $r_i$ is the disk inner
radius and $\beta$ is a parameter related to the angular momentum flux
(see, e.g., Spruit 2001). We rewrite the viscosity in terms of the
dimensionless $\alpha$ viscosity (Shakura and Sunyaev, 1973) as $\nu =
\alpha c_S H$, where
\be 
c_S = \left(\frac{kT}{\mu m_p}\right)^{1/2} = f \Omega r = f \left(\frac{G
    M_\star}{r}\right)^{1/2}\label{sound} 
\ee 
is the sound speed ($\mu\simeq 0.6$ is the mean atomic weight for a
gas of primordial composition, and $f= H/r$ is the disk aspect ratio).
We have used here the relation between the actual temperature in the
disk and the virial temperature,
\be T = f^2 T_{\rm{vir}} \simeq 1.4\times 10^5 f^2 \frac{M_\star}{10 M_\odot}\left(\frac{r}{10^3
    R_\odot}\right)^{-1} \ \rm{K}.\label{temp} \ee Combining eqs.
(\ref{surf}), (\ref{accrate}) and (\ref{sound}), we obtain an
expression for the midplane density \be \rho_0(r) = \frac{1}{6 \pi
  \alpha f^{3}} \frac{\dot{M}}{\sqrt{G M_\star r^3}}\left[1-\beta
  \left(\frac{r_i}{r}\right)^{1/2}\right].\label{dens} 
\ee
Furthermore, if the central star rotates at a rate $\Omega_\star <
\Omega_{\rm{Kep}}$, one finds that $\beta = 1$, and at large radii $r$
as compared to the disk inner radius $r_i$, the last term between the
square brackets above can be ignored.

We are now left with the task of evaluating the aspect ratio, $f$.
The few available models of disks around a forming Pop III star do not
really agree on the disk thickness (and overall structure). Tan and
McKee (2004) found a thin disk solution with $f\simlt 0.13$ up to
$r\sim 20$ AU whereas Mayer and Duschl (2005) show that, in their
``fiducial model'' with $\dot{M}=10^{-4} M_\odot/\rm{yr}$, $f\sim 0.4$
up to $r\sim 46$ AU, and grows to reach unity at larger radii.

\subsection{Magnetic field seeds in the disk}

Various models have been proposed for the creation of cosmological
magnetic fields in astrophysical plasmas in the early ages of the
post-recombination Universe. Most of them rely either on thermal
pressure effects (the so-called Biermann battery, Biermann 1950; for
applications in cosmology, see e.g. Kulsrud et al. 1997, Hanayama et
al. 2005) or on radiation drag (e.g. Harrison 1970, Langer, Puget and
Aghanim 2003, Langer, Aghanim and Puget 2005) operating on
cosmological distances. Those usually lead to relatively weak fields
on large scales, and one has then to rely on amplification by
contraction or dynamo and make additional assumptions (e.g. flux
freezing, geometry of collapse, etc.) in order to get interesting
values of fields in circumstellar disks.  A particularly effective 
turbulent   amplification
mechanism that operates in a dilute plasma such as the intracluster
medium or galactic halo appeals to a self-accelerating fluctuation dynamo
where anisotropic pressure gradients generate plasma instabilities that
result in a high  Reynolds number \cite{schek}.
%(Schekochihin and    Cowley 2006).
However, the same physical
processes are also at work in the disk itself, and comparatively high
fields may be directly created \emph{in situ}. We therefore do not
need to have recourse to preliminary magnetic seeds generation on
larger scales.

For a two component system, and with $m_e \ll \mu m_p$, combining the
ion and electron equations of motion leads to the generalized Ohm's
law
\be
\vec{E} = \frac{\vec{B}\times \vec{v}_e}{c} + \frac{\grad (n_e kT)}{en_e} + \frac{\vec{F}_{\rm{rad}}}{e},
\ee
the curl of which yields the induction equation
\be
\pd_t \vec{B} = \rot\vec{v}_e\times \vec{B} + c  \rot\frac{\grad(n_ekT)}{en_e} + \frac{\rot \vec{F}_{\rm{rad}}}{e}.
\ee
The first term on the right hand side above corresponds to the dynamo
amplification, and since we are interested here in the creation of
magnetic fields starting with $B = 0$, we will ignore this term in the
following. The second and last are the battery and radiation drag
terms respectively. Both can serve as sources to create a seed of
magnetic fields in the disk.
	
As is well known, the Biermann battery term is non zero only when the
temperature and electron density gradients are not parallel, according
to
\be
\pd_t \vec{B} = -\frac{ck}{e}\frac{\grad n_e \times \grad T}{n_e}.
\ee

The characteristic time scale is the rotation time $\Omega$, and the
electron density and temperature vary over scales at least of order
$H$ and $r$. Therefore, we expect the field created by the battery
mechanism to be of order
\bea
B_{\rm{batt}}&\sim& \frac{ck}{e} f^{-1}\frac{T}{\sqrt{GMr}}\nn\\
& \simeq& 3.93\times 10^{-12} f \left(\frac{M_\star}{10
    M_\odot}\right)^{1/2}\left(\frac{r}{10^3 R_\odot}\right)^{-3/2}\ \
\rm{G}.  
\eea 
Turbulence might actually create density gradients on smaller scales,
leading to higher field amplitudes, but by considering longer length
scales, we adopt the more conservative approach and keep the lower
estimate for $B$.

The value we obtained from the Biermann effect is just a factor $f$
smaller than that provided by radiation drag. Indeed, as has been
shown by Balbus (1993; see also Chuzhoy 2004) through momentum
conservation arguments, the typical amplitude we may expect is
\bea
B_{\rm{rad}}&\sim&\frac{\mu m_p c \Omega}{e}\nn\\
&\simeq& 3.93\times 10^{-12}\left(\frac{M_\star}{10 M_\odot}\right)^{1/2}\left(\frac{r}{10^3 R_\odot}\right)^{-3/2}\ \ \rm{G}
\eea
It so happens that this amplitude, $B_{\rm{rad}}$, corresponds to
values of the ions Larmor radius of order the local disk radius. This
means that the magnetic field is only weakly coupled to the ions.
Therefore we expect the advection of field lines by small scale
turbulence to be quite inefficient.

\subsection{MRI dynamo instability}
We now wish to derive a constraint on the mass of the central object
by requiring the magnetic field in the disk to exceed the minimal
value for the magneto-rotational instability (MRI) to operate and
drive a dynamo.

The first basic condition we consider is that the wavelength of
unstable modes must exceed the particle mean free path.  This
condition gives the minimum field for MRI.  Now in MRI (Balbus and
Hawley 1998), one expects that the dominant growth mode satisfies
$\lambda_{\rm{mi}}\approx v_A/\Omega,$ where $v_A$ is the Alfv\'en
frequency and $\Omega$ is the disk rotation rate. Larger scales become
unstable as the dynamo operates, and the maximum scale is given by the
disk scale height.  This gives the maximum field strength where the
dynamo saturates.  For a sustained field, that does not reverse sign
every rotation period, one requires a stratified disk, which allows
helical turbulence to be generated.  Helical turbulence results in
mean field amplification via the MRI dynamo and it is possible that
the resulting fields can attain the equipartition value (Brandenburg
2004).  Now the mean free path in an ionised gas is $9T^2k^2/(\pi n_e
e^4\ln \Lambda)$ (e.g., Lang 1999), and the minimum field condition is
thus
\be
\frac{v_A}{\Omega}> \frac{9}{\ln \Lambda}\frac{T^2k^2}{\pi n_e e^4},
\ee
or
\bea
B> B_{\rm{mfp}}&=& 1.75\times 10^{-9}\alpha^{1/2}f^{11/2}\frac{20}{\ln \Lambda} \left(\frac{M_\star}{10M_\odot}\right)^{11/4}\nn\\
&\times& \left(\frac{\dot{M}}{10^{-4}M_\odot/\rm{yr}}\right)^{-1/2} \left(\frac{r}{10^3 R_\odot}\right)^{-11/4} \rm{G}, 
\eea
where we assumed $\Omega = \Omega_{\rm{Kep}}$.  Notice that MRI may
occur even in collisionless plasmas, as shown by Quataert, Dorland and
Hammett (2002). Their study, however, does not account for possible
effects due to a finite Larmor radius, which we presumably cannot
ignore in our case where initially weak fields imply very large Larmor
radii.

The second condition we may consider is that MRI will be efficient in
amplifying magnetic fields only if the amplification time scale is
shorter than the diffusion time scale. For maximum instability scales,
the growth rate is of order $\Omega$, and the condition can be
rewritten as
\be
\frac{\eta}{\lambda_{\rm{mi}}^2 \Omega}  = \frac{\eta\Omega}{v_A^2}< 1
\ee
where $\eta = 10^{13}(\ln \Lambda/20) T^{-3/2}\ \rm{cm}^2/s$ is the
magnetic diffusivity. Assuming Keplerian rotation, this condition
translates to
\be
B^2> B_{\rm{mi}}^2 = 4\pi \eta \rho \sqrt{\frac{GM_\star}{r^3}}.
\ee
Using eqs. (\ref{temp}) and (\ref{dens}), we obtain
\bea
B> B_{\rm{mi}} &=& 4.93\times 10^{-8} \alpha^{-1/2} f^{-3} \left(\frac{\ln \Lambda}{20}\right)^{1/2}\left(\frac{M_\star}{10M_\odot}\right)^{-3/4}
\nn\\
&\times&\left(\frac{\dot{M}}{10^{-4}M_\odot/\rm{yr}}\right)^{1/2}\left(\frac{r}{10^3 R_\odot}\right)^{-3/4}\  \rm{G}.
\eea
As noticed by Tan and Blackman (2004) who obtained a similar result,
this value of the amplitude is quite high for initial conditions in
primordial star formation which starts in a protostellar environment
not necessarily magnetized beforehand.  However, it does not imply
that MRI could not be at work in a disk surrounding the progenitor of
a Population III star. Indeed, as already stressed, the minimum value
obtained above has been derived by taking into account effects of
magnetic diffusivity on scales of maximal instability, which are small
with respect to the disk height. This only implies that due to
diffusivity, MRI for small-wavelength disturbances is \emph{de facto}
inefficient, unless the amplitude of the field is already large.
Moreover, for standard values of the parameters, $B_{\rm{mi}}$ is way
above $B_{\rm{rad}}$. This means that $l_{\rm{mi}}$, the scale of
maximal instability, is much smaller than the ion Larmor radius, and
therefore, as noticed in the previous paragraph, MRI turbulence is
unable to amplify the fields on such small scales.%, contrary to the
%implicit assumtion of Tan and Blackman (2004).

Nevertheless, larger scales of order the disk (or Larmor) radius, may
still remain unstable, with the growth rate in that case being
proportional to the field strength itself (e.g. Balbus 1995), even if
magnetic fields are dissipated on small scales.  This brings us to the
third and strongest condition that must be satisfied. In a
differentially rotating, turbulent disk, the growth rate for MRI in
the weak field limit is $\gamma\sim v_A k,$ where $k$ is the vertical
wavenumber, where $v_A \ll\Omega/k$ and $\Omega$ is the local angular
velocity. The growth is stabilised by magnetic diffusivity, $\eta ,$
so that the dispersion relation is $\gamma +\eta k^2\approx v_A k$
(e.g. Kitchatinov and R\"udiger 2004). Therefore, for marginal
stability,
\be
v_A \simeq \eta \frac{\pi}{\lambda}
\ee
with $\lambda < H = f r$. This criterion translates readily into a minimum magnetic field
\be
B_{\rm{mri}} = 2\pi^{3/2}\eta \rho^{1/2}f^{-1} r^{-1}
\ee
which yields 
\bea
B_{\rm{mri}} &=& 3.87\times 10^{-15} \alpha^{-1/2} f^{-11/2}\frac{\ln \Lambda}{20} \left(\frac{\dot{M}}{10^{-4}M_\odot/\rm{yr}}\right)^{1/2}\nn\\ 
& & \times \left(\frac{M_\star}{10M_\odot}\right)^{-7/4}\left(\frac{r}{10^3 R_\odot}\right)^{-1/4}\ \ \rm{G}.
\eea
using again eqs. (\ref{temp}) and (\ref{dens}) above.

\subsection{Critical mass of the central object}
We require now that the minimum magnetic field for the MRI to work is
smaller than the field created in the disk. As we have seen, the
smallest suitable field amplitude, $B_{\rm{mri}}$, is given by the
marginal instability criterion applied to scales of order the disk
scale. At the same time, \emph{in situ} magnetic field generation is
likely to yield amplitudes of order $B_{\rm{rad}}$. This amplitude
must be higher than $B_{\rm{mri}}$, which happens as soon as
\bea
{M_\star} &>& 0.46  \alpha^{-2/9}f^{-22/9}\left(\frac{\ln \Lambda}{20}\right)^{4/9}\left(\frac{\dot{M}}{10^{-4}M_\odot/\rm{yr}}\right)^{2/9}\nn \\
&&\times
 \left(\frac{r}{10^3 R_\odot}\right)^{5/9} {M_\odot}.
\eea
The thickness of the disk has the most dramatic effect on the value of
the critical mass, whereas the actual value of the viscous $\alpha$
parameter plays a minor role.  If the disk is rather thin, with $f\sim
0.1$ for instance, then the mass spans a range of rather large values,
going from $128$ to $357 M_\odot$. However, in the absence of metals,
the properties of the primordial gas are likely to prevent the disk
from efficiently cooling, and we expect $f$ to be closer to unity. In
that case, taking $f\sim 0.4$ as obtained by Mayer and Duschl (2005),
the critical mass is much smaller, comprised between roughly $4.3$ and
$27.8 M_\odot$ for $\alpha$ ranging from $1$ to $0.01$.

\section{Discussion}
Magnetic fields have probably played a role in Primordial Star
Formation, even if starting from a medium free of magnetic fields. As
we have argued, radiation drag or thermal pressure effects are able to
generate magnetic seeds in the disk surrounding the central accreting
stellar progenitor. Initially, those seeds are weak too much for being
amplified by small scale turbulence, or even MRI effects on scales
comparable to the disk scale. However, as matter accretion proceeds,
the mass of the central object grows, and the gravitational potential
it creates deepens, increasing the rotation velocity of the disk.
Eventually, the rotation is fast enough so that long wavelength MRI
modes become unstable, as the minimum magnetic field for MRI becomes
smaller than the field generated in the disk. Depending on the
properties of the disk, this happens once the mass of the central
object reaches $4 - 28 M_\odot$.

Subsequently, MRI dynamo will be at work and will amplify the magnetic
field. The field can then rapidly reach dynamically important values,
and magnetically-driven ejection contributes to lower the effective
accretion efficiency. The actual model of magnetic winds is beyond the
scope of this article.  but this suggests that feedback is likely to
require magnetically-driven outflows, which could occur during Pop III
formation already when, as we argued above, the mass of the stellar
progenitor is rather small.  
 Note that, in their model of protostellar disk in primordial star
  formation, Tan and Blackman (2004) estimated the power of
  magnetically driven outflows. In their study, the outflow feedback
  effects reduce the star formation efficiency once the protostar
  reaches roughly 100 $M_\odot$.
 Incidentally, concurrent conclusions
  were reached by Machida et al.  (2006) who simulated the collapse of
  a magnetized primordial cloud in rigid rotation, and the subsequent
  formation of a Pop III star, within the ideal MHD approximation.
  Depending on the (high) initial value ($B_{\rm init} \gtrsim 10^{-9}
  G$ for an initial cloud density $n_{\rm c}\sim 10^3 {\rm cm}^{-3}$)
  of the magnetic field, their simulations show that magnetically
  driven jets develop and effectively reduce the accretion rate.
  Further numerical studies with higher resolution seem 
  necessary to address both the mass ejection rate and the
  amplification of magnetic fields in proto-stellar disks for
  initially weaker, even vanishing, magnetic seeds.  

We have
argued that the interplay between the two modes of star formation,
primordial, massive and conventional, involving all masses, is
controlled by $B$ and not by $Z$.  Effective feedback requires that of
order 10 percent of the gas accretion rate is channelled into star
formation, with an outflow rate that on the average must be of the
order of the net star formation rate. Globally, for the star forming
cloud, one expects that $\dot M_{\rm outflow} \sim \dot M_{\ast}\sim
0.1 \dot M_{\rm accretion} $, much as is found in nearby cases of
star-forming clouds.

Moreover, feedback is likely to be responsible for the turbulent
support in clouds that lowers the star-formation efficiency and helps
to generate the conventional IMF.  Thus the onset of cloud
fragmentation, due eventually but we have argued, not exclusively, to
the enhanced role of cooling, would allow field amplification, angular
momentum transfer, feedback and low mass star formation.

Alternatively, suppose we accept the hypothesis that the first stars
were massive objects.  To avoid the empirical objections discussed
above, one would have to argue that merging and coagulation of gas
clumps resulted in formation of predominantly very massive objects, of
characteristic mass $\simgt 1000\rm M_\odot,$ whose fate is to form
intermediate mass black holes with relatively low nucleosynthetic
yields associated with their collapse.  In this case, the seed fields
may come from jets and outflows driven by spin-up of turbulent
accretion disk dynamos as a consequence of accretion onto these
intermediate mass black holes.  One can argue that the first
generation of primordial clouds, which cooled predominantly via Lyman
alpha emission, preferentially formed IMBHs, since the associated
minihalos of mass $\sim 10^7\rm M_\odot$ are promising environments
for forming IMBHs in view of the high core accretion rates (Zhao and
Silk 2005). Accretion disks around the IMBHs provide promising sites
for MRI dynamos.

Alternatively, the jets may generate magnetic fields via the Weibel
instability, that although small, amounting to $\sim m_e/m_p$ of the
equipartition value, could still be useful as seeds (Wiersma and
Achterberg 2004).  The IMBH outflows allow one to generate larger seed
fields required for conventional star formation than inferred from the
Biermann battery.  The associated nucleosynthetic implications are
modest, because IMBHs have essentially zero yield. For any reasonable
IMF, there are of course associated Population III stars.

Consider for example the inference from primordial star simulations
that one star of mass $\sim 10^2 \rm M_\odot$ forms per pristine
pregalactic cloud of baryonic mass $\sim 10^7 \rm M_\odot$, a typical
galactic precursor not dissimilar in mass to dwarf satellite galaxies.
The resultant local enrichment would amount to of order $10^{-5}$ the
solar value, and would be diluted by a further factor that
corresponded to the fractional number of primordial star-forming
clouds that formed the eventual galaxy. However the associated
population of IMBHs could amount to as much as $\sim 10\%$ of the
current epoch stellar mass, if accretion onto these IMBHs also
provides a possible explanation of the NIR background excess (Madau
and Silk 2005).  If such IMBH-induced jets carry flux, produced by a
circum-black hole accretion disk dynamo, or drive jets that generate
magnetic fields vis the proton Weibel instability in the shock
interaction zone, then one might easily imagine seeding the clouds
with a magnetic field that more than sufficed to allow MRI to operate.

In summary, we have outlined two alternative pathways which dispense
with the need for Population III to consist exclusively of very
massive stars and for the transformation from first stars into
Population II to be determined exclusively by the rise in gas and/or
dust phase metallicity.  MRI-dynamo generated magnetic flux in the
protostellar accretion disk is the key to these possible scenarios.
The nature of the seed field is the major difference.  One option is
that protogalactic Biermann battery-seeded MRI dynamos provide
circumstellar disks with the necessary field strength to overcome high
accretion rates characteristic of primordial environments. This
allows, and indeed the mechanism itself requires, stars of mass below
$10-100\rm M_\odot$ to form.  In this case, there is no Population
III.  Another possibility is that the first objects (Population III is
a misnomer since no nucleosynthetic tracers remain) were very massive
($\simgt 1000\rm M_\odot$), and form IMBHs which expel
magnetic-flux-loaded jets that have been generated by dynamo activity
in the circum-IMBH disk. The MRI prostellar dynamos are seeded by
jet-induced instabilities, and again, the first stars are expected to
span a broad mass range. The IMF is a consequence of magnetic
feedback. Of course, it may vary with epoch and/or environment, but
the principal point we wish to emphasize is that it spans the
conventional range of stellar masses. One could then make the
transition from Primordial Star Formation to Current Epoch Star
Formation while the mean overall metallicity remained extremely low.

\section*{acknowledgments}
We thank J. Tan for stimulating comments. The work of M. Langer was
supported by a Marie Curie Individual Fellowship of the European
Community ``Human Potential'' programme (contract number
HPMF-CT-2003-02176).

\def\mnras{MNRAS}
\def\araa{ARAA}
\def\apj{ApJ}
\def\aj{AJ}
\def\pasp{PASP}
\def\apjl{ApJ}


\begin{thebibliography}{}
\bibitem{ABN}
Abel T., Bryan G.  and Norman M.,  2002, Science, 295, 93
\bibitem{Balbus1}
Balbus S., 1993, ApJ, 413, L137
\bibitem{Balbus2}
Balbus S., 1995, ApJ, 453, 380
\bibitem{BH}
Balbus S. and Hawley J., 1998, RMP, 70, 1
\bibitem{bapa}
Balbus S. and Papaloizou J., 1999, ApJ, 521, 650
\bibitem{BCG}
Bessell M., Christlieb N. and Gustafsson G., 2004, ApJL, 612, L61
\bibitem{battery}
Biermann L., 1950, Z.  Naturforsch, 5a, 65
\bibitem{Brandenburg}
Brandenburg A., 2004, astro-ph/0412366
\bibitem{BCL}
Bromm V.,  Coppi P. and Larson R.,  2002, ApJ, 564, 23
\bibitem{BL}
Bromm V.  and Loeb A.,  2003, Nature, 425, 812
\bibitem{Christlieb}
Christlieb N. et al., 2004, ApJ, 603, 708
\bibitem{Chuzhoy}
Chuzhoy L., 2004, MNRAS, 350, 761
\bibitem{Dietrich}
Dietrich M., Hamann F., Appenzeller I.  and Vestergaard M., 2003,
ApJ, 596, 817
\bibitem{Draine}
Draine B. T., Roberge W. G. and Dalgarno A., 1983, ApJ, 264, 485
\bibitem{Frebel}
Frebel A. et al., 2005, Nature, 434, 871
\bibitem{Gammie}
Gammie C. F., 2001, ApJ, 553, 174
\bibitem{Hanayama}
Hanayama H. et al., 2005, ApJ, 633, 941
\bibitem{Harrison}
Harrison E.R., 1970, MNRAS, 147, 279
\bibitem{KR}
Kitchatinov  L. L.  and R\"udiger G., 2004, A\&A, 424, 565
\bibitem{Klessen}
Klessen R. et al., 2004, astro-ph/0410371
\bibitem{Kulsrud}
Kulsrud R. M., Renyue C., Ostriker J. P. and Ryu D., 1997, ApJ, 480,
481
\bibitem{KB}
Kunz M. W. and Balbus S. A., 2004, MNRAS, 348, 355
\bibitem{Lang}
Lang K.R., Astrophysical Formulae (Springer, New York, 1999)
\bibitem{LPA1}
Langer M., Puget J.-L. and Aghanim N., 2003, Phys. Rev. D 67, 043505
\bibitem{LPA2}
Langer M., Aghanim N. and Puget J.-L., 2005, A\&A, 443, 367
\bibitem{LS}
Lizano S., and Shu F. H., 1989, ApJ, 342, 834
\bibitem{lori}
Lodato G. and Rice W.,2005, MNRAS, 358, 1489
\bibitem{MS}
Madau P.  and Silk J.,  2005, MNRAS, 359, L37
\bibitem{Machida}
Machida M. N., Omukai K., Matsumoto T. and Inutsuka S., 2006, astro-ph/0605146
\bibitem{MD} 
Mayer  M. and Duschl W., 2005, MNRAS, 356, 1
\bibitem{Mestel65}
Mestel L., 1965, QJRAS, 6, 161
\bibitem{Mestel}
Mestel L., 2003, \emph{Stellar magnetism}, Clarendon Press
(International series of monographs on physics), Oxford
\bibitem{NOP}
Norman M.,  O'Shea B. and Paschos P., 2004, ApJL, 601, L115
\bibitem{QDH}
Quataert E., Dorland W. and Hammett G.W., 2002,  ApJ, 577, 524
\bibitem{SMU}
Saigo K., Matsumoto T. and Umemura M., 2004, ApJL, 615, L65
\bibitem{alpha}
Shakura N. I. and Sunyaev R. A., 1973, A\&A, 24, 337
\bibitem{Shaye}
Schaye J. et al., 2003, ApJ, 596, 768
\bibitem{Schek}
Schekochihin, A.  and    Cowley, S., Phys. Plasmas, in press  (2006).
\bibitem{SLA}
Shu F. H., Li Z.-Y. and Allen A., 2004, ApJ, 60, 930 
\bibitem{SS1}
Spaans M. and Silk J., 2000, ApJ, 538, 115
\bibitem{SS2}
Spaans M. and Silk J., 2005, ApJ, 626, 644
\bibitem{Spruit}
Spruit H.C.,  2001, Accretion Disks, astro-ph/0003144
\bibitem{TB}
Tan J., and Blackman E.G., 2004, ApJ, 603, 401
\bibitem{TM}
Tan J. and McKee C., 2004, ApJ, 603, 383
\bibitem{UM1}
Umeda H. and Nomoto K., 2003, Nature, 422, 871
\bibitem{UM2}
Umeda H. and Nomoto K., 2005, ApJ, 619, 427
\bibitem{Wardle}
Wardle M., 1999, MNRAS, 307, 849
\bibitem{SW}
Salmeron R. and Wardle M., 2003, MNRAS, 345, 992
\bibitem{WA}
Wiersma J. and  Achterberg A., 2004, A\&A, 428, 365
\bibitem{ZS}
Zhao H. and Silk J., 2005, Phys. Rev. Lett., 95, 011301
\label{lastpage}

\end{thebibliography}
\end{document}